\documentclass[11pt,a4paper]{article}

\usepackage{jcappub}
\usepackage{amssymb}
\usepackage[numbers,sort&compress]{natbib}
\usepackage{amsmath,amsfonts,graphicx,epsfig}
\usepackage{bm}
\usepackage{hyperref}
\usepackage{cleveref}
\usepackage{xcolor}
\usepackage{amsmath,latexsym}
\usepackage{braket}
\usepackage[titletoc,toc]{appendix}

\def\bea{\begin{eqnarray}}
\def\eea{\end{eqnarray}}
\def\be{\begin{equation}}
\def\ee{\end{equation}}
\def\ba{\begin{array}}
\def\ea{\end{array}}

\def\P{{\mathcal P}}

\def\xip{{\xi}}
\def\f{{f_{\rm PBH}}}
\def\Ms{{\text{M}_{\odot}}}

\newcommand{\dd}{\mathrm{d}\baselineskip0.5ex}
\newcommand{\eqdot}{\, .}
\newcommand{\eqcom}{\, , \quad}
\newcommand{\erfc}{\text{erfc}}

\def\xM{{x_{\rm max}}}
\def\xm{{x_{\rm min}}}

\title{LIGO/Virgo black holes and dark matter: The effect of spatial clustering}

\author{Vicente Atal,}
\author{Albert Sanglas}
\author{and Nikolaos Triantafyllou}

\affiliation{Departament de F\'isica Qu\`antica i Astrof\'isica, i  Institut  de  Ci\`encies  del  Cosmos, Universitat de Barcelona, Mart\'i i Franqu\`es 1, 08028 Barcelona, Spain.}

\emailAdd{vicente.atal@icc.ub.edu}
\emailAdd{asanglas@icc.ub.edu}
\emailAdd{nitriant@icc.ub.edu}

\abstract{We discuss the effect of clustering for the determination of the merger rate of binary black holes in the LIGO/Virgo mass range. While for a Poissonian initial distribution, and assuming isolated binaries, the allowed fraction of Primordial Black Holes (PBHs) to dark matter (DM) is a few percent,  we show that this bound can be relaxed if PBHs are clustered. More precisely we show that for large clustering the merger rate  can drop with increasing fraction of PBHs, introducing a degeneracy in the parameters of the theory consistent with a given present merger rate, and allowing all the DM to be in the form of stellar mass PBHs.  This degeneracy can however be broken by looking at the evolution of the merger rate  with redshift. For a simple clustering model that we consider, we show that the LIGO/Virgo projected sensitivity can disentangle, through the observation of a stochastic background of gravitational waves, different clustered distributions having the same present merger rate.}

\begin{document}
\maketitle
\flushbottom

\section{Introduction} 

The merger rate of binary black holes observed by LIGO provides one of the strongest constraints for the presence of Primordial Black Holes (PBH) of masses $\mathcal{O}(1-100)$ $\Ms$ \cite{Abbott:2017vtc,Bird:2016dcv,Clesse:2016vqa,Sasaki:2016jop}. Assuming that PBHs follow an initial Poissonian spatial distribution  -and that binaries are isolated objects (see \cite{Raidal:2018bbj,Jedamzik:2020ypm,Jedamzik:2020omx} for caveats)- it has been found that PBH binaries merging today are mainly formed in the radiation dominated Universe \cite{Nakamura:1997sm,Sasaki:2016jop}, and that they can account for a small fraction of the dark matter (DM) in this range of masses, of the order of a few percent \cite{Sasaki:2016jop,Ali-Haimoud:2017rtz}\footnote{PBHs of masses around $10^{-16}-10^{-11} \, \Ms$ can still form all of the dark matter under these assumptions  \cite{Katz:2018zrn} (see \cite{Carr:2020gox,Green:2020jor} for recent reviews on the constraints for the presence of PBHs in all the mass ranges).}. A related constraint, although weaker in this mass range, comes from the non-observation of a stochastic background of gravitational waves (GWs), generated from past binary mergers \cite{TheLIGOScientific:2016wyq, Wang:2016ana}.

Poissonian spatial distributions for PBHs arise if the density field is a Gaussian random field with a spiky power spectrum \cite{Ali-Haimoud:2018dau,Desjacques:2018wuu,Ballesteros:2018swv} (for earlier discussions, see \cite{Chisholm:2005vm}). On the contrary, if PBH arise from non-Gaussian perturbations, then their distribution will not be Poissonian. In particular, a coupling between small and long wavelengths of the density perturbations, or the modulation of the density field by a secondary field,  can result in a clustered spatial distribution \cite{Tada:2015noa,Young:2015kda,Belotsky:2018wph,Suyama:2019cst,Young:2019gfc}. Its effect on the merger rate  of late and early time binaries has been estimated in \cite{Clesse:2016vqa,Raidal:2017mfl,Clesse:2017bsw,Ballesteros:2018swv,Vaskonen:2019jpv,Ding:2019tjk}, and on the stochastic background of gravitational waves in \cite{Bringmann:2018mxj}. Other effects can alter the merger rate, such as enhanced large scales perturbations \cite{Eroshenko:2016hmn,Ali-Haimoud:2017rtz,Garriga:2019vqu}, three-body \cite{Jedamzik:2020ypm,Jedamzik:2020omx} and  many-body interactions \cite{Raidal:2018bbj} and PBH mass accretion \cite{DeLuca:2020qqa}

In this paper we re-examine the question of how an initial clustering of PBH can affect the bounds on their abundance coming  from present and past merger rate  of binaries. Clustering enhances the local density of PBHs and so it has been usually found that the allowed fraction of PBH to DM, $\f$, is smaller than that of a Poissonian initial distribution (see e.g. \cite{Bringmann:2018mxj}). As we will show, this is only the case in the limit in which $\f$ is very low. For larger abundances, the present merger rate drops, and larger values of $\f$ are allowed with respect to the Poissonian case. This degeneracy, i.e. the fact that the same merger rate  can be achieved with different values of $\f$,  can be  disentangled either by directly measuring the merger rate  as a function of redshift (possible with future experiments like Cosmic Explorer \cite{Evans:2016mbw} or Einstein Telescope \cite{Punturo:2010zz}), or by measuring the stochastic GW background created by past mergers (possible with current experiments like LIGO/Virgo \cite{Abbott:2017vtc}).

We will show explicitly how this works in a specific model for the clustering, in which the reduced local threshold for gravitational collapse, $\nu(x)$, is a  local parameter linearly related to a secondary field $\psi(x)$. This simple model will allow us to analytically compute the correlation function $\xip(r)$ encoding the properties of the clustering. 
In particular, we will show how a comparison between the present merger rate and the stochastic GW background can help us disentangle the initial distribution of PBHs.\\

We will begin in Section \ref{sec:1} by calculating the PBH abundance and correlation function for a simple model of non-Gaussianity, inducing clustering. We will then compute in Section \ref{sec:mergerrate} its effect on the present and past merger rates, and show in Section \ref{sec:3} how LIGO/Virgo capabilities in measuring the stochastic background of gravitational waves can determine the initial distribution of PBHs. Throughout this paper we use natural units, $G=c=1$.

\section{The PBH abundance and correlation function}\label{sec:1}

Clustering appears if the local properties of the overdensity field $\delta(x)$ are space dependent. The probability of forming a PBH can then be modeled as depending on a ``secondary" field $\psi(x)$. This  effect might either come from an actual field different from the overdensity field, or from a long wavelength modulation of the overdensity field itself, resulting from a self-coupling of long and short scales as happens e.g. in local models of non-Gaussianity \cite{Komatsu:2001rj}. In both cases, the field $\psi(x)$ acts as a long wavelength modulation of the small scale perturbations $\delta(x)$, inducing a local change on the variance $\sigma_{\delta}(x)\equiv\sigma_{\delta}(\psi(x)) $, and/or on the threshold for collapse into a BH, $\delta_c(x)\equiv \delta_c(\psi(x))) $.  Assuming that both fields are independent and that $\delta(x)$ is locally Gaussian, then the probability for forming a BH depends on the local reduced threshold $\nu(x)\equiv\delta_c(x)/\sigma_\delta(x)$ in the following way \footnote{For simplicity we compute the abundance of PBHs and the correlation function that encodes the clustering using the Press-Schechter formalism. More accurate criteria for the formation of PBH can be obtained from the statistics of peaks \cite{Bardeen:1985tr}, although we do not expect qualitatively differences with the results obtained here. The statistics of peaks has been recently revisited to count for peaks in the so-called compaction function \cite{Germani:2019zez}, which is the object that controls the critical collapse for the formation of a BH  \cite{Shibata:1999zs}.}
\be
P_1^{\text{local}}=\dfrac{1}{2}\erfc \left(\dfrac{\nu(x)}{\sqrt{2}}\right) \ .
\ee
This means that local changes in $\delta_c$ or $\sigma_\delta$ are in practice indistinguishable. The total probability for a given region to form a BH, that we denote $P_1$, is obtained by integrating over the configurations of the field $\psi(x)$, that we assume to be a Gaussian random field. Similarly we can define $P_2^{\text{local}}(r)$, the joint probability of having two black holes at a distance $r$, given that the local reduced thresholds at $x_1$ and $x_2$ are given by $\nu(x_1)$ and $\nu(x_2)$. This is given by \cite{Owen2} 
\begin{equation} \label{p2after}
\begin{aligned}
P_2^{\text{local}}(r) &= \dfrac{1}{4}\erfc\left(\dfrac{\nu_1}{\sqrt{2}}\right) + \dfrac{1}{4}\erfc\left(\dfrac{\nu_2}{\sqrt{2}}\right) + \dfrac{\text{sgn}\left(\nu_1\right)\text{sgn}\left(\nu_2\right) - 1}{4} \\
&- T\left(\nu_1, \dfrac{\nu_2-\omega_\delta(r) \nu_1}{\nu_1\sqrt{1-\omega_\delta^2(r)}}\right) - T\left(\nu_2, \dfrac{\nu_1-\omega_\delta(r) \nu_2}{\nu_2\sqrt{1-\omega_\delta^2(r)}}\right) \ ,
\end{aligned}
\end{equation}
where $\omega_\delta(r) =\langle \delta(x_1) \delta(x_2)\rangle / \langle \delta(0)^2\rangle$ is the reduced correlation function of $\delta(x)$, $\nu_i=\nu(x_i)$ and $T(z,a)$ is the Owen T-function \cite{Owen1956}
\begin{equation}
T(z,a) \equiv \dfrac{1}{2\pi} \int_0^a \dd{t}\dfrac{e^{-\frac{(1+t^2)z^2}{2}}}{1+t^2} \ .
\end{equation}
The total $P_2(r)$ can be found by integrating this expression over the fields $\psi_1\equiv\psi(x_1)$ and $\psi_2\equiv\psi(x_2)$ inducing the spatial dependence on $\nu_i(x)$.

The clustering of PBHs can be encoded in the $N$-point correlation functions $\xi^{(N)}(r)$, that measure the excess probability, relative to an uncorrelated distribution, of finding  $N-1$ black holes at distances $r_1,...,r_{N-1}$ from a BH at $r=0$. This is then given by
\be
\xi^{(N)}(r_1,...,r_{N-1})=\frac{P_{N}(r_1,...,r_{N-1})}{P_1^N}-1 \ .
\ee
As we will see, the merger rate  of BHs depends on all the $N$-point correlation functions (as shown in \eqref{eq:En_corr} of Appendix \ref{app:nearest_neighbour}). However in some cases  all that information is contained in the 2-point correlation function $\xi^{(2)}(r)\equiv \xi(r)$. Whenever this is not possible, we will show that the use of $\xi(r)$ can nevertheless provide useful insights into the qualitative behaviour of the mergers.

In the following we briefly discuss the Gaussian case.

\subsection{Gaussian case}
The Gaussian case is recovered if we turn-off the field $\psi(x)$. In this case $\nu_1 = \nu_2 = \nu_g$, where $\nu_g\equiv\delta_{c,g}/\sigma_\delta$ is space independent (and $\delta_{c,g}$ is the threshold for collapse in the Gaussian case). Then we get
\begin{equation}\label{eq:p1p2_gauss}
P_1=\dfrac{1}{2}\erfc\left(\dfrac{\nu_g}{\sqrt{2}}\right) \quad\text{and}\quad P_2(r) = \dfrac{1}{2}\erfc\left(\dfrac{\nu_g}{\sqrt{2}}\right) - 2 T\left(\nu_g, \sqrt{\dfrac{1- \omega_\delta(r)}{1+\omega_\delta(r)}}\right) \ ,
\end{equation}
and the $2$-point correlation function $\xi(r)$ is given by \footnote{Some properties of Owen T-functions can be found in \cite{Owen2}.}
\begin{equation}
\xi(r) = 2\dfrac{T\left(\nu_g,1\right)-T\left(\nu_g,\sqrt{\dfrac{1-\omega_\delta(r)}{1+\omega_\delta(r)}}\right)}{P_1^2} \ .
\end{equation}
This expression is exact, valid for any $\nu_g$ and $\omega_\delta$. 
Simpler expressions  can be obtained in the regime for which $\nu$ is large or small \cite{Kaiser:1984sw,Ali-Haimoud:2018dau}.  For example, for large $\nu$, which is the relevant limit for PBH formation, we can make use of the expansion
\be
T(\nu,a)\sim \frac{1}{4}\erfc \left(\frac{\nu}{\sqrt{2}}\right)-\frac{1}{2\pi}\frac{e^{(-1+a^2)\nu^2/2}}{ \nu^2 a (1+a^2)} + \mathcal{O}\left(\frac{e^{-\nu^2}}{\nu^4}\right) \ ,
\ee
and then $\xi(r)$ is given by \cite{Ali-Haimoud:2018dau}
\begin{equation}\label{eq:equi_gaus}
1 + \xi(r) \sim \dfrac{\left(1+\omega_\delta(r)\right)^{\frac{3}{2}}}{\left(1-\omega_\delta(r)\right)^{\frac{1}{2}}} \exp\left(\nu_g^2\dfrac{\omega_\delta(r)}{1+\omega_\delta(r)}\right) \ .
\end{equation}

In the following we will present a simple model for which the correlation function in the non-Gaussian case can be computed.

\subsection{A simple model for clustering}

Here we consider a simple model of clustering were the parameter $\nu(x)$ is linearly related to the secondary field $\psi(x)$ as\footnote{This relation can be seen as the first order term of a Taylor expansion in the field $\psi(x)$ around $\psi=0$. Let us note that if $\psi=0$ corresponds to an extremum of $\nu(\psi)$, then the expansion would start at second order in $\psi(x)$. It might then be interesting to study generalizations of this model, even if in principle we would expect similar qualitative effects on the merger rate. We thank Jaume Garriga for pointing this out. }
\be\label{eq:linearmodel}
\nu(x)=\nu_g\left(1+\beta\psi(x)\right) \ ,
\ee
\noindent where $\beta$ denotes the strength of the coupling between $\psi(x)$ and $\delta(x)$. With this simple model we can solve for $P_1$ and $P_2(r)$ exactly, getting (see Appendix \ref{app:Nth_point_dist})
\begin{align}\label{totalabundance}
P_1 =& \dfrac{1}{2} \erfc\left[\dfrac{\nu_g}{\sqrt{2}}\dfrac{1}{\sqrt{1+\alpha^2}}\right]  \ , \\ \label{totalabundance2}
P_2(r) =& \dfrac{1}{2}\erfc\left(\dfrac{\nu_g}{\sqrt{2}}\dfrac{1}{\sqrt{1+\alpha^2}}\right) - 2 T\left(\dfrac{\nu_g}{\sqrt{2}}\dfrac{1}{\sqrt{1+\alpha^2}}, \sqrt{\dfrac{1- \bar{\omega}}{1+\bar{\omega}}}\right) \ .
\end{align}
Here
\be
\alpha \equiv \delta_{c,g} \beta \left(\dfrac{\sigma_\psi}{\sigma_\delta}\right) \quad \text{and}\quad \bar{\omega}(r) = \dfrac{\omega_\delta(r)+\alpha^2\omega_\psi(r)}{1+\alpha^2}  \ ,
\ee
\noindent where $\sigma_\psi$ and $\sigma_\delta$ stands for the variance of the long and short wavelength perturbations respectively. Let us notice that the effective coupling is determined by $\alpha$, which not only takes into account the coupling between both fields (given by $\beta$), but is also sensitive to the relative amplitude of the variances. By comparing \eqref{totalabundance} and \eqref{totalabundance2} with the expressions found for the Gaussian case \eqref{eq:p1p2_gauss}, we see that this model is equivalent to a Gaussian model with the replacements
\be
\nu_g\rightarrow \bar{\nu}=\frac{\nu_g}{\sqrt{{1+\alpha^2}}} \quad\text{and}\quad \omega_g \rightarrow \bar{\omega} \ .
\ee 
This is actually true for all $N$-point probabilities, as shown in the Appendix \ref{app:Nth_point_dist}. In particular, this implies that in this model the total abundance is amplified with respect to the Gaussian case. 
For concreteness let us choose a two-point correlation function for $\omega_\delta$ and $\omega_\psi$ as given from a peaked power spectrum at both scales. In particular we choose \footnote{This is the power spectrum for the density fluctuation evaluated at the time when the small scales perturbations $\delta$ enters the horizon.}
\begin{equation}\label{eq:PS_deltas}
 \mathcal{P}_i(k) = \sigma_i^2 k_i \delta(k-k_i) 
\end{equation}
where $i=(\delta,\psi)$. Here the short mode $k_\delta$ contributes to the formation of the PBHs and the long mode $k_\psi$ modulates the amplitude of the short one. As both $\psi(x)$ and $\delta(x)$ are Gaussian random fields, their two-point correlation function $\xi_i(r)$ is given by \cite{Bardeen:1985tr} 
\begin{align}
\xi_i(r) &= \int d \ln k \, \P_i(k) \frac{\sin (k_i r)}{k_i r} \ .
\end{align}
We then have that
\begin{equation}
\omega_i(r) = \dfrac{\sin\left(k_i r\right)}{k_i r} \ .
\end{equation}
\begin{figure}[tpb]
\includegraphics[width=0.5\linewidth]{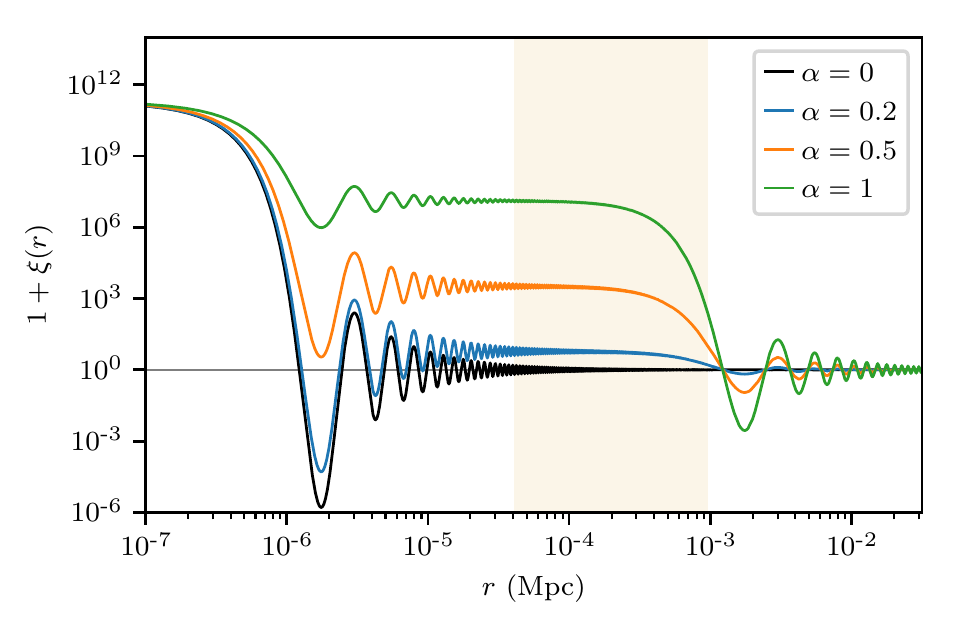}
\includegraphics[width=0.5\linewidth]{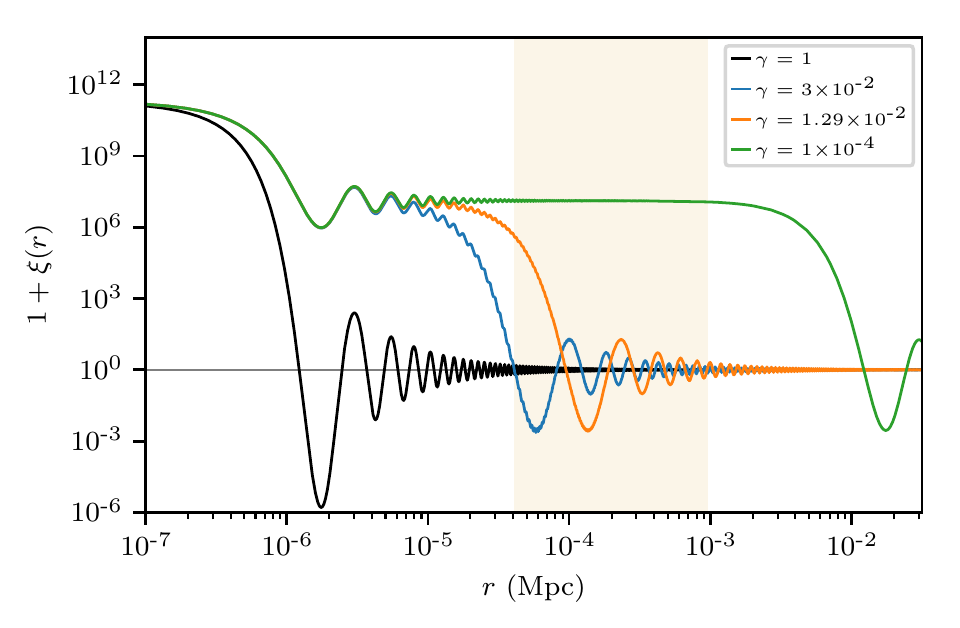}
\caption{The two-point correlation function $\xip(r)$ for the linear model \eqref{eq:linearmodel} and for power spectra given by the delta functions \eqref{eq:PS_deltas}. \textit{Left}) The correlation function $\xip(r)$ as we vary the strength of the coupling $\alpha$ between $\delta(x)$ and $\psi(x)$. For $\alpha\ll1$, the Gaussian case is recovered. Here we fix $\gamma=10^{-3}$. \textit{Right}) The correlation function $\xip(r)$ as we vary the hierarchy between the short and large wavelengths, given by $\gamma \equiv k_\psi/k_\delta$. We fix $\alpha=1$. In both left and right panels we choose $\bar{\nu}=6.8$ and $k_{\delta} \simeq 2.5{\times}10^6 \text{ Mpc}^{-1}$, which corresponds to a present PBH abundance of $f_{\text{PBH}} = 10^{-3}$ for BHs of masses $M = 30 \Ms$. The orange region $(x_{\text{min}},x_{\text{max}})$ indicates the scale of binaries that merge today $(t \sim 14 \text{Gyr})$. The first drop in the correlation function corresponds to the size of the BH, $R_{\text{BH}} \sim k_\delta^{-1}$, so the region below this point reflects the autocorrelation of $\delta(r)$. The second drop corresponds to the typical scale of the secondary field $R_{\text{cl}}\sim k_\psi^{-1}$ and it defines the clustering length of the PBHs. Beyond $R_{\text{cl}}$ the correlation is effectively zero so the distribution becomes Poissonian.} \label{fig:correlationDelta}
\end{figure}
In Figure \ref{fig:correlationDelta} we show how $\xip(r)$ varies as a function of the strength of the coupling $\alpha$ (left panel), the relative scales between the short and long wavelengths, $\gamma\equiv k_{\psi}/k_{\delta}$, and the fraction of DM in form of PBHs, $\f \equiv \Omega_{\rm DM}/\Omega_{\rm PBH}$. Let us note that for the Gaussian case, corresponding to $\alpha=0$ or $\gamma=1$,  all the $N$-point correlation functions are zero for $r>R_{\rm BH}$, meaning that the distribution of BHs is Poissonian \cite{Ali-Haimoud:2018dau,Ballesteros:2018swv}. On the other hand as can be seen in Figure \ref{fig:correlationDelta}, in the non-Gaussian regime there is a region for $r>R_{\rm BH}$ where the $2$-point correlation is constant and possibly large. This plateau is a consequence of the nearly constant correlation function $\bar{\omega}$ induced by the long wavelength perturbation. For $N$ BHs within this region, it is possible to calculate their $N$-point correlation function for  large thresholds. These are given by  \cite{Ruben1964AnAE}
\begin{equation}\label{eq:equi_1}
1 + \xi^{(N)}(\bar{\omega},\bar{\nu}) \sim \dfrac{\left(1+(N-1)\bar{\omega}\right)^{N-\frac{1}{2}}}{\left(1-\bar{\omega}\right)^{\frac{N-1}{2}}} \exp\left(\dfrac{N \bar{\nu}^2}{2}\dfrac{(N-1)\bar{\omega}}{\left(1+(N-1)\bar{\omega}\right)}\right) \ , 
\end{equation}
which is valid when $\bar{\omega} \in (-1/(N-1),1)$. For $N=2$ we have
\begin{equation}\label{eq:equi_2}
1 + \xi^{(2)}(\bar{\omega},\bar{\nu}) \sim \dfrac{\left(1+\bar{\omega}\right)^{\frac{3}{2}}}{\left(1-\bar{\omega}\right)^{\frac{1}{2}}} \exp\left(\bar{\nu}^2\dfrac{\bar{\omega}}{\left(1+\bar{\omega}\right)}\right) \ .
\end{equation}
In general not all the BHs determining the properties of the binary  will necessarily be within this region, since the $N$-point correlation eventually vanishes for $r>1/k_\psi$. This estimate provides then an upper bound for the amplitude of their correlation.

\section{Merger rate}\label{sec:mergerrate}
Now we consider the effect of the clustering on the merger rate  of binaries. When the torque of the binary (preventing a head-on collision) is provided by a third BH, we need to consider the probability density function  to find the nearest BH at distance $x$ and the next-to-nearest BH at distance $y$ from a reference BH at $r=0$. We call this distribution $Q(x,y)$ and in general it takes the following form (see Appendix \ref{app:nearest_neighbour})
\begin{equation}\label{eq:mergingrate_full}
Q(x,y) = 16\pi^2 n^2 x^2 y^2 G_0(x) G_1(y) \exp\left[-4\pi n\left(\int_0^x G_0(z) z^2\dd{z} +\int_x^y G_1(z) z^2\dd{z}\right)\right] \Theta(y-x) \ ,
\end{equation} 
with $n$ the  comoving number density of BHs and where $G_m(r)$, the $m$-particle conditional pair correlation function, refers to the conditional probability of finding a BH at radius $r$ given that there is one at $r=0$ and $m$ additional BHs in the interior of the region of radius $r$ \cite{truskett1998density}. The functions $G_m(r)$ depend on all the $N$-point correlation functions and so in general they are hard to determine. However in some cases they take simple forms. For a Poissonian distribution, the presence of a BH at any given position is independent on the absence or presence of BHs at any other position implying $G_m=1$. Then, for a Poissonian distribution $Q(x,y)$ takes the following form (see e.g. \cite{Nakamura:1997sm}) 
\be\label{eq:mergingrate_1}
Q(x,y)=16\pi^2x^2y^2n^2\exp\left[-4\pi \int_{R_\text{BH}}^y n z^2\dd z\right]\Theta(y-x) \ .
\ee
Here $R_{\rm BH}$ is the radius of the BH at\footnote{As shown in Figure \ref{fig:correlationDelta}, for $r<R_{\rm BH}$, the function $\xi(r)$ measures the autocorrelation of $\delta(r)$.} $r=0$. Under certain conditions, a non-Poissonian distribution can also be written as in \eqref{eq:mergingrate_1}, provided that the comoving number density is promoted to a local density
\be\label{eq:local_density}
n\rightarrow n(r)= n\, g_2(r) \quad \textrm{with}\quad g_2(r)\equiv 1+\xi(r)   \ .
\ee
In particular, if the $N$-point correlation functions $\xi^{(N)}$ satisfy
\be\label{eq:sep}
1+\xi^{(N)}(r_1,...,r_{N-1})=\prod_{i=1}^{N-1}(1+\xi(r_i)) \ ,
\ee
then $G_m(r)=g_2(r)$ for all $m$ and $Q(x,y)$ takes the form \cite{Ballesteros:2018swv} (see also Appendix \ref{app:nearest_neighbour})
\be\label{eq:mergingrate_clus}
Q(x,y)=16\pi^2x^2y^2 n(x) n(y)\exp\left[-4\pi \int_{R_\text{BH}}^y n (z) z^2\dd z\right]\Theta(y-x) \ .
\ee
This is for example the case for biased Gaussian distributions, as our model, with a constant and small correlation function $\xi^{(N)}$ \cite{Jensen:1986ws}. In the case in which the correlation functions $\xi^{(N)}$ are larger than expected from the separability condition (\ref{eq:sep}) we expect to have 
\be
G_n(r) \geq g_2(r) \ .
\ee
For some distributions this inequality can be shown to hold explicitly \cite{torquato1990,torquato2008point,torquato2015}. Using the bounds found in e.g. \cite{torquato1986,torquato1990}, it can be shown that for small radius $G_0(r)\simeq g_2(r)$. For larger radius, and because the probability of finding a void decreases, we expect $G_0$ to be an increasing function of $r$. This would then imply that $G_0(r) \geq g_2(r)$ at all relevant scales (we also expect the same to happen for $G_1(r)$). Then the expression \eqref{eq:mergingrate_clus} becomes either an upper or a lower bound, depending on whether  the linear or the exponential term in \eqref{eq:mergingrate_full} dominates. In the rest of the paper, we test whether \eqref{eq:sep} holds using the $N$-correlation function as given by  \eqref{eq:equi_1}, for $N$ up to the expected number of BHs in the volume where the correlation is non-vanishing.\\

In the following we will compute the merger rate as given by the simple expression \eqref{eq:mergingrate_clus}, taking into consideration that this expression will be an upper or lower bound for the true merger rate for large correlations, as explained above.

In order to find the merger rate  at a given time $t$, we need to integrate $Q(x,y)$ over the positions $(x,y)$. Binaries that merge today were initially separated by a comoving distance $x$ in the interval ($\xm$, $\xM$). The distance $\xM$ is the maximum distance such that a pair of BHs can form a binary system, and can be found by imposing that the mass of the binary system is larger than the background mass within a volume whose radius is the initial separation of the binary \cite{Sasaki:2016jop,Raidal:2018bbj}. This radius is maximised at matter-radiation equality and then $\xM$ is given by\footnote{Slightly different estimates for $\xM$ are obtained depending on whether the volume is defined in cartesian or spherical coordinates.}
\be\label{eq:xM}
\xM\simeq \left(\frac{M}{\rho_{eq}}\right)^{1/3} \ ,
\ee
where $M=m_1+m_2$ is the total mass of the binary, $\rho_{eq}$ is the background energy density at matter-radiation equality. The distance $\xm$ is the distance below which a binary with any orbital parameter would have already merged. The time $t$ for an orbit to collapse with dimensionless angular momentum\footnote{More precisely $j=c\,(x/y)^3$, where $c$ is a factor $\mathcal{O}(1)$. We choose $c=1$.} is given by \cite{Peters:1964zz}
\be\label{eq:peters}
t=\frac{3}{85}\frac{r_x^4}{\eta M^3}j^7 \quad \textrm{with} \quad j=(x/y)^3 \ ,
\ee
where $r_x$ is the semi-major axis of the binary and $\eta\equiv m_1m_2/M^2$ is the symmetric mass ratio.  Then, for a given semi-major axis, the longest possible time for a binary to merge is if they have initially a circular orbit ($j=1$). The semi-major axis $r_x$ is proportional to the physical distance at the time the binary decouples from the Hubble flow. We then need to find the scale factor $a_{\rm dec}$ at which the condition
\be
M = \rho(a) r^3    
\ee
is satisfied, where $\rho(a)$ is the total energy density at radiation domination. Using $a_{\rm eq}=1$, we find
\be
r_x= \left(\frac{x}{\xM}\right)^3 x \ ,
\ee
and so
\be
\xm=\left(\frac{85\, \eta \, M^3 \,t}{3\, x^4_{\rm max}}\right)^{1/16}\xM \ .
\ee
For each position $x$ within $(\xm,~\xM)$ there is a corresponding $y$ such that the merging time is $t$, as given by Eq. \eqref{eq:peters}. For $t=t_0$ the present age of the Universe, and $m_1=m_2=30\, \Ms $, $\xm\simeq 4\times 10^{-5} \text{ Mpc}$ and $\xM\simeq 9.6\times 10^{-4} \text{ Mpc}$.   
By integrating $Q(x,y)$ over this interval we find the total merger rate  per unit time at a given time $t$. In order to find the merger rate  per unit time and volume element, we multiply by the total density of PBHs, $\bar{n}$. This is found by integrating $n(r)$ in \eqref{eq:local_density} over a Hubble patch and dividing by the total volume. As the radius of the Hubble patch under consideration is much bigger than $R_{\rm cl}\sim k_{\psi}^{-1}$, the contribution of $\xi(r)$ to the average number density of PBH is negligible, and so $\bar{n}\simeq n$. By using \eqref{eq:xM}  the number density is related to $\f$ by
\be
\bar{n}=\frac{\f}{\xM^3} \ .
\ee
Then, the merger rate  at a given time $t$ is 
\be\label{eq:mergerrate_2}
\frac{\dd R}{\dd t}=\frac{\bar{n}}{2}\int_{x_{\rm min}}^{x_{\rm max}} Q(x,y(x,t))\Big|\frac{\dd y}{\dd t}(t,x)\Big|dx \ ,
\ee
where the factor $1/2$ avoids overcounting the binaries.
\begin{figure}[tpb]
\includegraphics[width=0.5\linewidth]{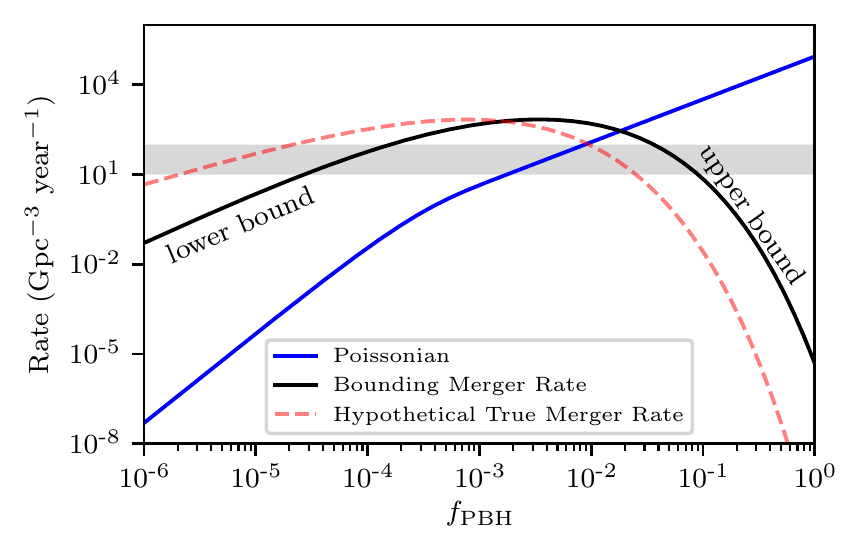}
\includegraphics[width=0.5\linewidth]{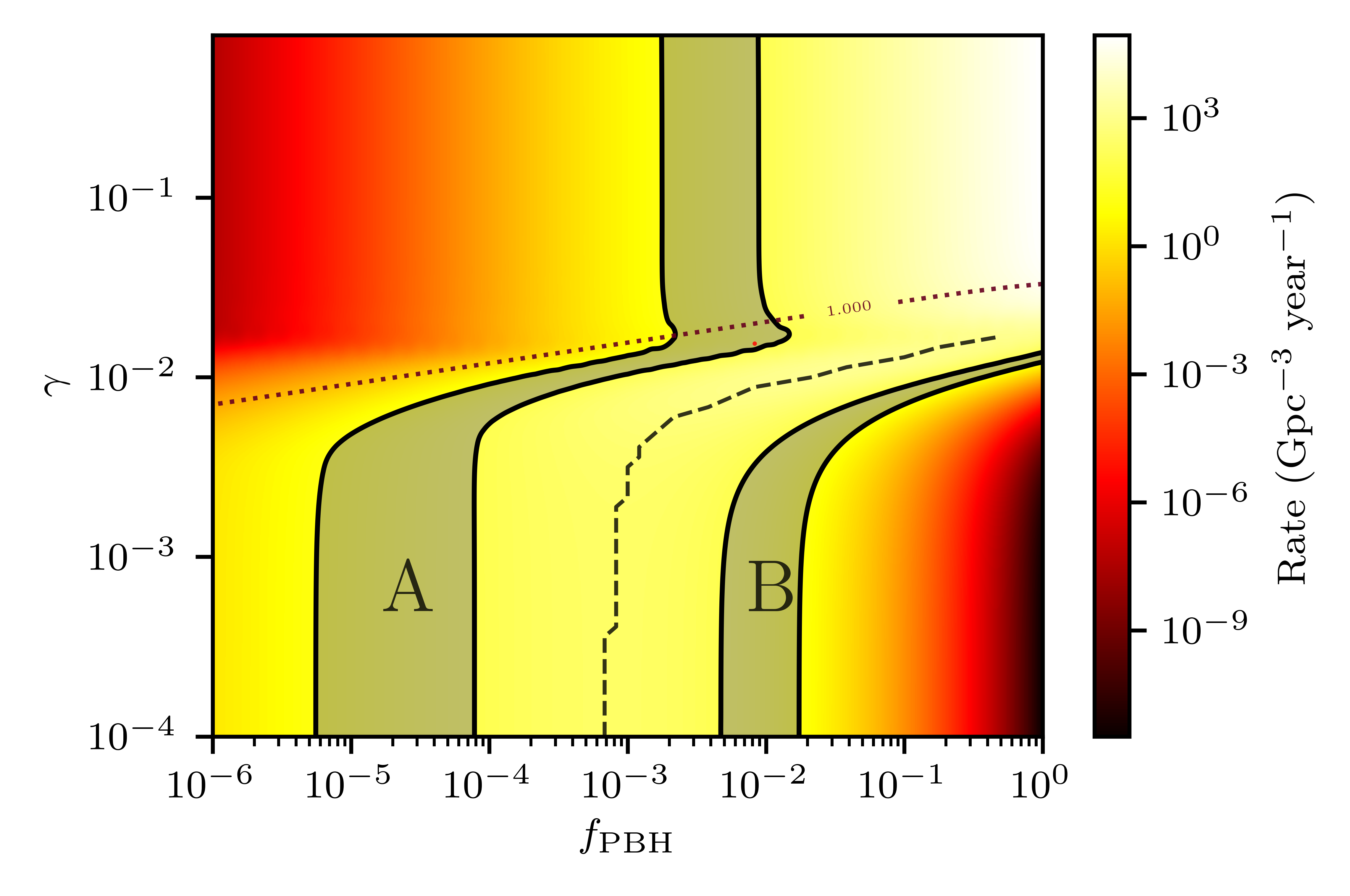}
\caption{\textit{Left}) For large clustering, $\gamma\ll1$, the  merger rate  in Eq. \eqref{eq:mergingrate_clus} (in black) becomes either an upper bound or lower bound of the unknown true merger rate  (depicted here in red), depending on whether the linear or the exponential term in \eqref{eq:mergingrate_full} dominates. For the Poissonian case (blue), the linear term always dominates. \textit{Right}) Present merger rate  from \eqref{eq:mergingrate_clus} for a monochromatic mass spectrum with $m=30\, \Ms$, as a function of $\f$ and $\gamma$ for $\alpha=1$. In the grey band we show the present rate of binary mergers as determined by LIGO/Virgo observations. We distinguish two branches, branch A which is linked to the Poissonian case ($\gamma=1$), and branch B that appears as a result of the drop in the merger rate  for large $\f$ and small $\gamma$. Above the dotted line the condition \eqref{eq:sep} holds, and the true merger rate  is accurately described by \eqref{eq:mergingrate_clus}. Below that line, the merger rate  depicted here is either an upper or lower bound of the true merger rate, depending on whether this rate is an increasing or decreasing function of $\f$ (as seen in the left panel). This limit is shown as a dashed line, where we show the position of the maximum of the merger rate. From here we deduce that the positions of branch A and B below the dotted line should be displaced towards smaller values of $\f$. This implies that, for a given $\gamma$, the merger rate  consistent with LIGO/Virgo will happen for smaller $\f$ than shown here.}
\label{fig:rates}
\end{figure}
In Figure \ref{fig:rates} we show the merger rates for the linear model \eqref{eq:linearmodel}, with the power spectra given by \eqref{eq:PS_deltas}.  We fix $\alpha=1$ and show the merger rate as a function of $\f$ and $\gamma\equiv k_{\psi}/k_{\delta}$. The Poissonian case corresponds to $\gamma=1$ (in such case there is no long wavelength modulation). From Figure \ref{fig:rates} we see that while for the Poissonian case the merger rate  increases monotonically with $\f$, for the clustered distribution ($\gamma\ll1$) the rate increases with increasing $\f$ only until a certain value of $\f$. For $\f\gtrsim 10^{-3}$ and $\gamma\lesssim 10^{-2}$, the merger rate  decreases with increasing $\f$. The drop is due to the fact that the exponential term in \eqref{eq:mergingrate_clus} dominates, which is never the case in the Poissonian case (if we would consider an unphysical $\f>1$ we would eventually also notice the exponential drop in this case).  

\begin{figure}[tpb]
\centering
\includegraphics[width=0.5\linewidth]{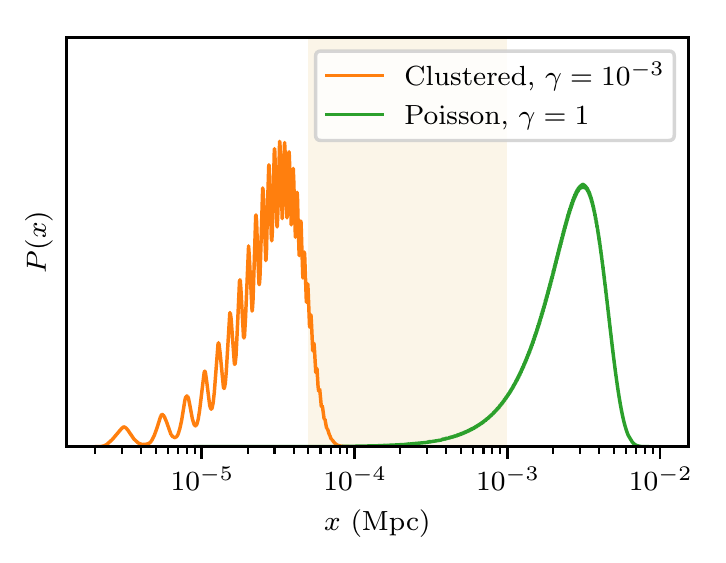}
\caption{Distribution of the comoving distance of the nearest BH in a Poissonian and in a clustered distribution of BHs, for parameters giving the same present rate ($(\gamma=1,\f=4.6\times 10^{-3})$ in the Poissonian case and $(\gamma=10^{-3}, \f=2.2\times 10^{-5})$ for the clustered distribution). The distributions are normalized so that they can be easily compared in the same figure. For a Poissonian distribution, most of BHs are separated by a distance $x>\xM$. For a clustered distribution, most of BHs are separated by a distance $x<\xm$. BHs at a distance in the interval $(\xm,~\xM)$, shown in grey, can merge by today (provided the third BH provides the right amount of torque).}\label{fig:xdistr}
\end{figure}

\begin{figure}[tpb]
\begin{center}
\includegraphics[width=0.7\linewidth]{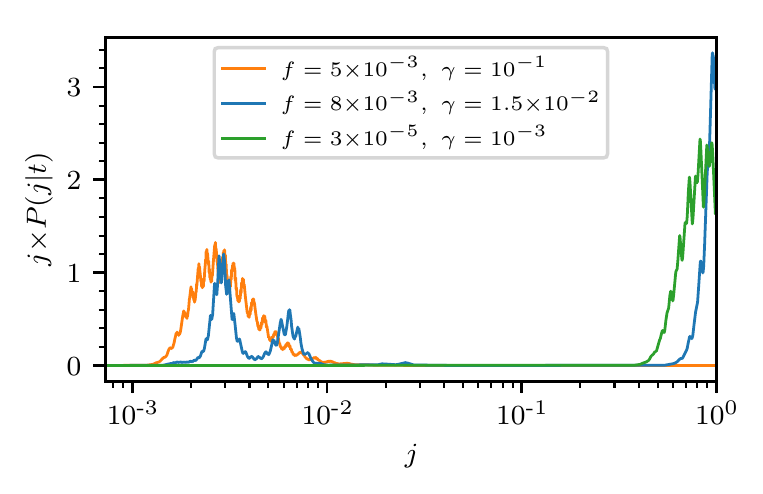}
\caption{Distribution of initial angular momenta for binaries merging today. Here we see three cases in branch A in which most binaries are eccentric (corresponding to a Poissonian distribution, in orange), circular (corresponding to a fully clustered distribution, in green) and mixed (corresponding to the intermediate case, in blue).}\label{fig:JsDistr}
\end{center}
\end{figure}

To understand the reason behind the drop in the merger rate, we recall that binaries merging today were initially separated by a distance $x$ inside the interval ($x_{\rm min}$, $x_{\rm max}$), as depicted in Figure \ref{fig:xdistr}. In the Poissonian case, BHs have a mean separation $\bar{x}$ much larger than $\xM$. Few of them would have a separation $x \lesssim x_{\rm max}\ll\bar{x}$, then forming a binary. These binaries merge today if their orbits were initially very eccentric (otherwise, their merging time is too large). As $\f$ increases, the typical distance between two BHs diminishes  and then it is more likely for them to have a separation $x \lesssim x_{\rm max}$. Then the merger rate increases.
The picture changes for the clustered distribution. For small $\gamma$, the typical distance to the nearest BH is smaller than $x_{\rm min}$. That is, binaries merging today are, contrary to the Poissonian case, rare binaries separated by a distance $x \gtrsim x_{\rm min}\gg\bar{x}$. These will merge today if their initial orbits are circular (otherwise their merging time is too short). 
As $\f$ increases, the typical distance between two BHs becomes even smaller, and so it is more rare to have binaries separated by a distance larger than $x_{\rm min}$. This explains the drop of the merger rate for $f>10^{-3}$ in the small $\gamma$ region\footnote{As we have already said, this drop in the merger rate  would also be visible in the Poissonian case if we would allow $\f$ to be much larger that unity.  In that hypothetical case, the typical distance of a binaries goes from being much larger than $\xM$, to be much smaller than $\xm$.}.
In Figure \ref{fig:xdistr} we show the typical distance of two BHs in the Poissonian and in the clustered regimes. In an intermediate regime there are two local maxima of the distribution. That behaviour is better seen by looking at the angular momentum of the binaries, that we show in Figure \ref{fig:JsDistr}.

The fact that the merger rate can drop as we increase $\f$ for clustered distributions  means that the rate observed by LIGO/Virgo is, for small and constant $\gamma$, consistent with two different values of $\f$\footnote{For very small $\f$ and $\gamma$, the rate still increases with $\f$, since, while the area under the curve in Figure \ref{fig:xdistr} is more or less constant with $\f$, there is a prefactor proportional to the total abundance of PBHs in Eq. \eqref{eq:mergerrate_2} that dominates the estimation of the total rate.}. 
This is the origin of the two branches of parameter space consistent with LIGO/Virgo that we see in Figure \ref{fig:rates}. One is connected to the Poissonian case ($\gamma=1$), and we call  it branch A.  We call branch B  the one resulting from the decrease in the rate at large clustering ($\gamma<10^{-2}$) and large $\f$.
In branch A and in the Poissonian case ($\gamma=1$), as we previously explained, most binaries merging today were initially very eccentric. As $\gamma$ decreases, and the probability to form a BH increases at small radii, a new population of BH appears with mean distance smaller than $\xm$. At some point as we move in branch A these two populations of binaries coexist, and we have a mixed population of binaries.
\begin{figure}[tpb]
\begin{center}
\includegraphics[width=0.7\linewidth]{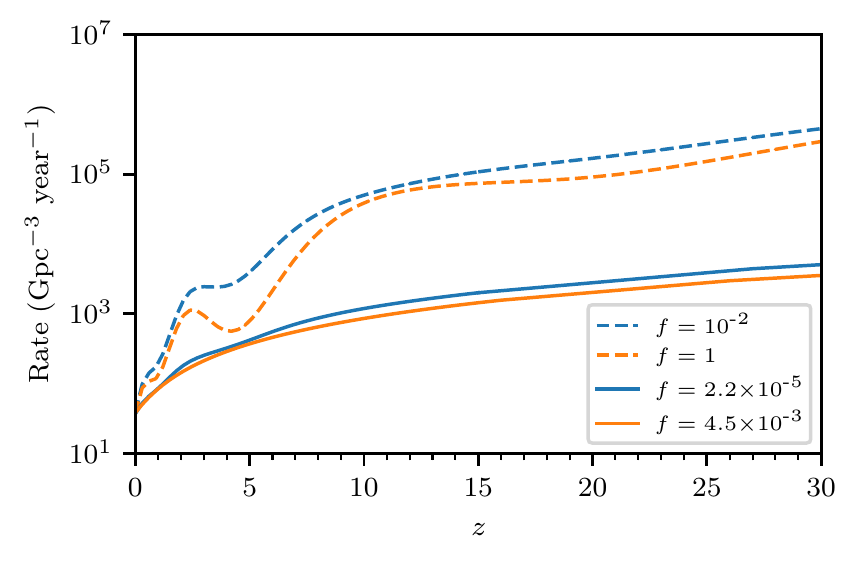}
\caption{Merger rate as a function of redshift $z$ for two different families of parameters. All of them have the same present merger rate, consistent with LIGO/Virgo, but  different values for $\f$. Solid lines corresponds to parameters in the branch A (($f=2.2\times10^{-5}$, $\gamma= 10^{-3})$ and ($f=4.5\times10^{-3}$, $\gamma=10^{-1}$)), and dashed lines corresponds to parameters in branch B (($f=1, \gamma=1.3\times 10^{-2}$) and ($f=10^{-2}, \gamma=10^{-3}$)). Note that, as we explained in the previous section, because we expect branch A and B to be displaced towards smaller $\f$ below the dotted line in Figure \ref{fig:rates} these curves should correspond to larger values of $\gamma$ than the ones quoted here. For example for $\f=1$, we expect $\gamma>1.3\times 10^{-2}$.}\label{fig:RatevsZeta}
\end{center}
\end{figure}
In order to assess the relative abundances of both excentric and circular orbit populations we need to determine the probability distribution of $j$  for   binaries merging at time $t$, $P(j|t)$, which can be related to $Q(x,y)$ as
\begin{align}
P(j|t)=S(t)^{-1} Q(x(j,t),y(j,t))\Big|\frac{\partial (x,y) }{\partial (j,t) }\Big| \ ,
\end{align}
where $x(j,t)$ and $y(j,t)$ can be found from Eq. \eqref{eq:peters} and $S(t)$ is a normalization factor. Notice that for a given merging time, the allowed separation of binaries $(\xm,~\xM)$ translates into possible values for $j$ in the interval ($j_{\rm min},1)$ with  $j_{\rm min}=(\xm/\xM)^{16/7}$. In Figure \ref{fig:JsDistr} we show the distribution of angular momenta for three cases having the same present merger rate  ($R\simeq 50$ Gpc$^{-3}$ yr$^{-1}$). These three cases follow a mostly Poissonian, clustered or mixed distribution of BHs. In the mixed case, there are two populations of circular and eccentric binaries contributing equally to the present merger rate.
The degeneracy between branch A and B is broken if we consider the merging history. In Figure \ref{fig:RatevsZeta} we show the merger rate  as a function of redshift, for four different parameters for which the present merger rate  is the same (two of them belong to branch A - solid lines- and two belong to branch B - dashed lines). While the merging history can then help us disentangle whether binaries come from branch A or B, the differences within each branch are less noticeable, in particular for the case of branch A. 
In principle there is a range of parameters for which the present merger rate  can be explained with $\f=1$, around $\gamma\simeq 0.01$ in Figure \ref{fig:rates}. We should however be cautious  since at large $\f$ most of the binaries are disrupted under the influence of others PBHs \cite{Raidal:2018bbj}, thus yielding smaller present merger rates\footnote{Let us note however that for slightly larger $\gamma$, a merger rate  larger by many orders of magnitude can be achieved. Thus it is possible that there is a $\gamma>0.01$ for which the merger rate  is consistent with $\f=1$, even if most of the mergers are disrupted at early times.}. A quantitative estimation on how this effect changes the rate is however only possible by the use of numerical simulations.

A direct detection of the merger rate  as a function of redshift  will of course first  contribute in determining whether these binaries are of primordial origin or not. For astrophysical binaries, the merger rate drops for $z>2$ and then dies off. These different histories might be directly disentagled with more events in the LIGO/Virgo channel \cite{Fishbach:2018edt} and with future experiments like the Cosmic Explorer \cite{Evans:2016mbw} or Einstein Telescope \cite{Punturo:2010zz} (see e.g. \cite{Chen:2019irf}). At last, let us note that the merger rate in branch B is several orders of magnitude larger that of branch A. For such large rates, the gravitational waves created by the binaries would be strongly lensed by other PBHs, leading to signatures that might explain some features of the LIGO/Virgo events \cite{Broadhurst:2018saj,Diego:2019rzc}.

The integrated effect of the merging history can also be deduced by looking at the stochastic background of gravitational waves, which we discuss in more detail in the following section. As we will show, when LIGO/Virgo acquires full capability, a detection of the stochastic background will make it possible to distinguish between these different merging histories.

\section{The Stochastic Background of Binary Mergers}\label{sec:3}

The energy released by binaries that have already merged contribute to a stochastic background of gravitational waves. The energy density of the stochastic background $\Omega_{\rm GW}$ can be expressed in terms of the critical density $\rho_c$ as
\be
\Omega_{\rm GW}\equiv \frac{1}{\rho_c}\frac{d \rho_{\rm GW}}{d \log \nu}
\ee
where $\rho_{\rm GW}$ is the energy density at a given frequency $\nu$. The contribution coming from early formed binaries can then be expressed as (see e.g. \cite{Wang:2016ana})
\be
\Omega_{\rm GW}=\frac{\nu}{\rho_c H_0}\int_{0}^{z_*}\frac{R_{\rm PBH}(z)}{(1+z)E(z)}\frac{d E_{\rm GW}}{d \nu_s}(\nu_s)dz
\ee
\noindent where $ d E_{\rm GW}/d \nu_s$ is the GW energy spectrum of the merger and $\nu_s$ is the frequency in the source frame, related to the observed frequency as $\nu_s=(1+z)\nu$. The function $E(z)\equiv H(z)/H_0=[\Omega_r(1 +z)^4+ \Omega_m(1 +z)^3+ \Omega_\Lambda]^{1/2}$.
Black holes of $m \sim 10^2\, \Ms$ were formed at $z_{*}\sim 10^{10}$, and so this is the maximum redshift at to which we possibly integrate the relation above (even though the integrand stops contributing much earlier).

\begin{figure}[tpb]
\begin{center}
\includegraphics[width=0.8\linewidth]{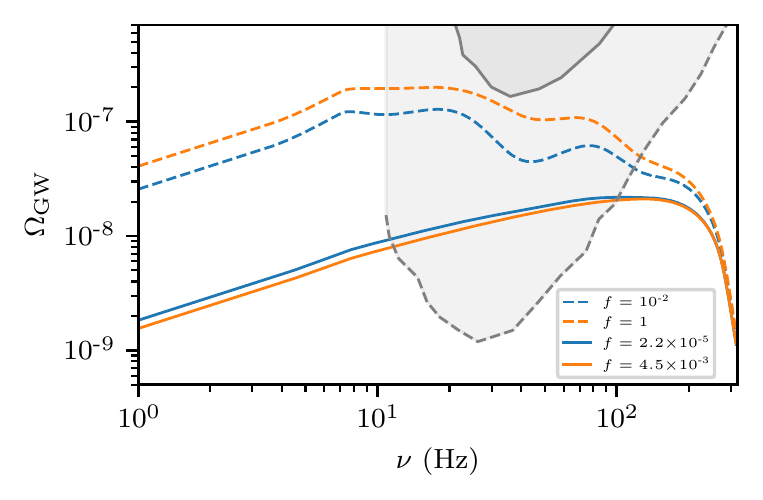}
\caption{Stochastic background of gravitational waves. We choose the same parameters as in Figure \ref{fig:RatevsZeta}, having the same present merger rate. The solid grey indicates the current LIGO/Virgo bounds whereas the dashed line indicates                                      the projected final sensitivity \cite{TheLIGOScientific:2016dpb}.}\label{fig:SBGW}
\end{center}
\end{figure}
As can be seen from Figure \ref{fig:SBGW}, under the hypothesis that all mergers are of primordial origin, LIGO/Virgo will be able to disentangle  the initial orbits of the binaries.

The energy released as GWs can be deduced from its waveform. This has been modeled for the inspiral, merger, and ringdown phases of BHs binary mergers, and fitted through numerical simulations \cite{Ajith:2007kx,Ajith:2009bn}.
For non-precessing binaries, as it is the case for solar mass PBHs \cite{Mirbabayi:2019uph,DeLuca:2019buf,DeLuca:2020bjf}, it takes the following form (see also \cite{Zhu:2011bd,Wang:2019kaf})

\be\label{eq:dEdnu}
  \frac{d E_{\rm GW}}{d\nu_s}(\nu_s) = \frac{\pi^{2/3}M_c^{5/3}}{3}
  \begin{cases}
    \nu_s^{-1/3} & \text{for } \nu_s < \nu_1 \\
    \omega_1\nu_s^{2/3} & \text{for } \nu_1 \leq \nu_s < \nu_2 \\
    \omega_2\frac{\sigma^4\nu_s^2}{\left(\sigma^2+4\left(\nu_s-\nu_2\right)^2\right)^2} & \text{for } \nu_2 \leq \nu_s < \nu_3 \\
    0 & \text{for } \nu_3 \leq \nu_s
    \end{cases}
\ee
\noindent where $\nu_i\equiv(\nu_1,\nu_2,\sigma,\nu_3)=(a_i \eta^2+b_i \eta + c_i)/(\pi M)$,  $M=m_1+m_2$ is the total mass, $M_c$ is the chirp  mass ($M_c^{5/3}=m_1m_2M^{-1/3}$),  and $\eta=m_1m_2 M^{-2}$ is the symmetric mass ratio. The parameters $a_i$, $b_i$ and $c_i$ can be found in \cite{Ajith:2007kx}, and ($\omega_1, \omega_2$) are chosen such that the spectrum is continuous. In the template \eqref{eq:dEdnu} the three regimes corresponds, towards larger frequencies, to the inspiral, merging and ringdown phases, and for $30\, \Ms$ they correspond to $\nu_i=(135,271,79,387) \text{ Hz}$.

In Figure \ref{fig:SBGW} we show the stochastic background of gravitational waves for a set of parameters producing the same present merger rate. We take $(\Omega_r, \Omega_\Lambda,\Omega_m)$ as determined by the Planck satellite \cite{Ade:2015xua}. On the one hand, let us note that the shape of the stochastic background of gravitational waves coming from astrophysical binaries is qualitatively very similar  to the one coming from branch A and from a Poissonian distribution of PBHs \cite{Chen:2018rzo,Safarzadeh:2020qru}, and thus a very careful comparison should be done in order to discriminate between the two.  On the other hand, the signal from branch B features oscillations  -residuals of the oscillations in the merger rate- that might act as a smoking gun for this type of distribution.    

\section{Discussion and Conclusions}

We have presented a model for initial clustering of PBHs and computed the merger rates of binary black holes. We have shown that because of the clustering the merger rate  can decrease as the fraction of black holes increases, inducing a degeneracy in the value of $\f$ for which a certain merger rate  is obtained and opening the possibility that all DM is in the form of PBHs of stellar mass.
We have showed that the detection of the stochastic background (within the projected sensitivity of LIGO/Virgo) should be able to break this degeneracy,  determining the initial distribution of PBHs.

While in this paper we have discussed the constraints coming from the present and past merger rates, there are other constraints that  put bounds on the abundance of PBHs at these scales. Depending on the mass function of the PBHs, more observables related to binary mergers can be used to confront with LIGO/Virgo data, such as the mass ratios, total mass and chirp mass \cite{Gow:2019pok}. Additionally, other observables not related to binaries can be used, such as distortions in the Cosmic Microwave Background (CMB) \cite{Ali-Haimoud:2016mbv,Poulin:2017bwe}, gravitational lensing of Type IA Supernova \cite{Zumalacarregui:2017qqd}, pulsar timings \cite{Schutz:2016khr}, and the survival of star clusters \cite{Brandt:2016aco}. In general we expect these constraints to be alleviated by the presence of clustered distributions (see e.g. \cite{Clesse:2016vqa,Garcia-Bellido:2017fdg,Clesse:2020ghq}), however a quantitative analysis taking into account the full space of parameters that we have considered here is still lacking.


Our analysis is based on binary orbits induced by a third BH. For large values of $\f$, as well as highly clustered distributions, this assumption might not hold. We thus expect some changes in the quantitative results for some of the parameters of the theory, when N-body effects are considered. Moreover, we used an estimate for the merger rate  that assumes the separability condition Eq. \eqref{eq:sep}, and we identified the regions of parameter space where this estimate is accurate, or provides a lower or upper bound of the true merger rate. This is sufficient for having a qualitative understanding of the possible merger histories at large clustering. A quantitative estimate can be obtained by calculating explicitly the probability of finding nearest neighbours at a given position, by applying e.g. the analytic techniques of \cite{torquato1986,torquato1990,truskett1998density}. These issues can also be tackled  by using numerical simulations to determine the initial distribution of BHs and by looking at their time evolution (see e.g. \cite{Trashorras:2020mwn} for recent considerations on these problems).  These lines of research will be further pursued in future work. Let us note that both of these effects make the merger rate  drop for larger abundances, and thus both contribute in opening the window for having all the DM as stellar PBHs.

\section*{Acknowledgments}

We thank Jaume Garriga for enlightening discussions and Salvatore Torquato for a stimulating correspondence. We also thank Jose M. Diego for comments on the manuscript. This work has been partially supported by FPA2016-76005-C2-2-P, MDM-2014-0369 of ICCUB (Unidad de Excelencia Maria de Maeztu), and AGAUR2017SGR-754. A.S is supported by an APIF grant from Universitat de Barcelona. N.T. is supported by an INPhINIT grant from la Caixa Foundation (ID 100010434) code LCF/BQ/IN17/11620034. This project has also received funding from the European Unions Horizon 2020 research and innovation programme under the Marie Sklodowska-Curie grant agreement No. 713673.

\section*{Appendix}
\addcontentsline{toc}{section}{Appendices}
\renewcommand{\thesubsection}{\Alph{subsection}}
\numberwithin{equation}{subsection}

\subsection{N-th point distribution of PBHs}\label{app:Nth_point_dist}
In this Appendix we compute the probability of having $N$ primordial black holes at the points $x_1,\dots , x_N$ for the case where the local critical threshold for collapse follows a linear relation with the secondary field $\psi$
\begin{equation}
\nu(x_i) = \nu_g \left(1 + \beta \psi(x_i)\right) = \nu_g + \alpha \mu(x_i) \eqcom
\end{equation}
where we have defined $\alpha  \equiv \nu_g \beta \sigma_\psi$ and $\mu(x_i) = \psi(x_i)/\sigma_\psi$. We denote by $P_N^{\text{local}}(\bm{x})$ the probability of having $N$ primordial black holes at points $x_1,\dots , x_N$ given that the local threshold for collapse has a value $\nu(x_1),\dots,\nu(x_N)$. We denote this probability by $P_N^{\text{local}}(x_1,\dots,x_N)$. By using threshold statistics\footnote{The main result of this Appendix also applies to peak theory.} we write this probability as 
\begin{equation}
P_N^{\text{local}}(x_1,\dots,x_N) = \int_{\nu(x_N)}^{\infty}\dd{\eta_N} \cdots \int_{\nu(x_1)}^{\infty}\dd{\eta_1} P^{(N)}\left(\eta,\Omega_\delta\right) \eqdot
\end{equation} 
Here we introduced the quantity $P^{(N)}(\eta,\Omega_\delta)$, the joint probability density of the $\delta$-field with correlation matrix $\Omega_\delta^{ij} = \omega_\delta(r_{ij})$ if $i\neq j$ and $1$ if $i = j$. Here $r_{ij}=|x_i-x_j|$. We also defined the vector $\eta = (\eta_1,\dots,\eta_N)$, $\eta_i = \delta(x_i)/\sigma_\delta$. The explicit form for $P^{(N)}(\eta,\Omega_\delta)$ is 
\begin{equation}
P^{(N)}\left(\eta,\Omega_\delta\right) = \dfrac{1}{\left(2\pi\right)^{N/2} \left(\det \Omega_\delta\right)^{1/2}}\exp\left(-\dfrac{1}{2} \eta^T \Omega_\delta^{-1} \eta\right) \eqdot
\end{equation}
The quantity $P_N^{\text{local}}(x_1,\dots,x_N)$ is a conditional probability. To obtain the total probability of finding PBH at the points $x_1,\dots,x_N$ we need to integrate over the configurations of the secondary field $\psi(x)$. Therefore, we have
\begin{equation}\label{pnBefore}
P_N(x_1,\dots,x_N) = \int_{-\infty}^{\infty}\dd{\mu_N} \dots \int_{-\infty}^{\infty}\dd{\mu_1} P^{(N)}\left(\mu,\Omega_\psi\right) P_N^{\text{local}}(x_1,\dots,x_N) \eqcom
\end{equation}
where $\mu$ and $\Omega_\psi$ are analogous to $\eta$ and $\Omega_\delta$ for the secondary field $\psi$. By using the change of variables
\begin{equation}
\tilde{\eta} = \dfrac{\eta-\alpha\mu}{\sqrt{1+\alpha^2}} \eqcom \tilde{\nu} = \dfrac{\alpha \mu}{\sqrt{1+\alpha^2}} 
\end{equation}
and defining the new covariance matrices $\tilde{\Omega}_\delta = (1+\alpha^2)^{-1}\Omega_\delta$ and $\tilde{\Omega}_\psi = \alpha^2(1+\alpha^2)^{-1}\Omega_\psi$ we can write Eq. \eqref{pnBefore} as
\begin{equation}
P_N(x_1,\dots,x_N) = \int_{-\infty}^{\infty}\dfrac{\dd^N{\tilde{\mu}}}{\sqrt{\left(2\pi\right)^N \det \tilde{\Omega}_\psi}} \int_{\bar{\nu}}^{\infty} \dfrac{\dd^N{\tilde{\eta}}}{\sqrt{\left(2\pi\right)^N \det \tilde{\Omega}_\delta}} e^{- \frac{1}{2} (\tilde{\eta}+\tilde{\mu})^T \tilde{\Omega}_\delta^{-1} (\tilde{\eta}+\tilde{\mu}) - \frac{1}{2} \tilde{\mu}^T \tilde{\Omega}_\psi^{-1}\tilde{\mu}} 
\end{equation}
where we used the notation $\dd^N{z} = \dd{z_N} \cdots \dd{z_1}$ and we have defined $\bar{\nu} = \nu_g / \sqrt{1+\alpha^2}$. By using the following known result for gaussian integrals
\begin{equation}\label{matrixidentity}
\dfrac{1}{\left(2\pi\right)^{N}} \int_{-\infty}^{\infty}\dd^N z e^{- i y^T z - \frac{1}{2} z^T M z} = \dfrac{1}{\sqrt{\left(2\pi\right)^N \det M}} e^{-\frac{1}{2} y^T M^{-1} y } \eqcom
\end{equation}
we can write
\begin{equation}
P_N(x_1,\dots,x_N) = \int_{-\infty}^{\infty}\dd^N{\tilde{\mu}} \int_{\bar{\nu}}^{\infty} \dd^N{\tilde{\eta}} \int_{-\infty}^{\infty}\dfrac{\dd^N \eta'}{\left(2\pi\right)^N} \int_{-\infty}^{\infty}\dfrac{\dd^N \mu'}{\left(2\pi\right)^N} e^{- i \tilde{\mu}^T \mu' - i (\tilde{\eta}+\tilde{\mu})^T \eta' - \frac{1}{2} \mu'^T \tilde{\Omega}_\psi \mu -\frac{1}{2} \eta'^T \tilde{\Omega}_\delta \eta } 
\end{equation}
We can first integrate over $\tilde{\mu}$ giving us a factor of $(2\pi)^N \delta^N (\mu' + \eta')$ allowing us to perform automatically the integral on $\mu'$. Then, we are left with 
\begin{equation}
P_N(x_1,\dots,x_N) = \int_{\bar{\nu}}^{\infty} \dd^N{\tilde{\eta}} \int_{\infty}^{\infty} \dfrac{\dd^N\eta'}{\left(2\pi\right)^N} e^{-i \tilde{\eta}^T \eta' - \frac{1}{2}\eta' \left(\tilde{\Omega}_\delta + \tilde{\Omega}_\psi\right) \eta } \eqdot
\end{equation}
By defining 
\begin{equation}
\bar{\Omega} \equiv \tilde{\Omega}_\delta + \tilde{\Omega}_\psi = \dfrac{\Omega_\delta + \alpha^2 \Omega_\psi}{1+\alpha^2}
\end{equation}
and using again the identity \eqref{matrixidentity} we end up with
\begin{equation}
P_N(x_1,\dots,x_N) = \int_{\bar{\nu}}^{\infty} \dd{\tilde{\eta}_N} \dots \int_{\bar{\nu}}^{\infty} \dd{\tilde{\eta}_1} P^{(N)}\left( \tilde{\eta},\bar{\Omega}\right) \eqdot
\end{equation}
But this is just the probability of finding $N$ primordial black holes at the points $x_1,\dots,x_N$ if the overdensity field was a \textit{single} gaussian field with correlation matrix $\bar{\Omega}$ and the threshold for the collapse was $\bar{\nu}$. There exist ``closed forms`` for the some values of $N$. 

For $P_1(x_1)$ we have
\begin{equation}
P_1(x_1) = \dfrac{1}{2}\erfc\left(\dfrac{\bar{\nu}}{\sqrt{2}}\right) \eqcom
\end{equation}  
and for $P_2(x_1,x_2)$ we have \cite{Owen1956}
\begin{equation}
P_2(x_1,x_2) = \dfrac{1}{2}\erfc\left(\dfrac{\bar{\nu}}{\sqrt{2}}\right) - 2 T\left(\bar{\nu},\sqrt{\dfrac{1-\bar{\omega}(r)}{1+\bar{\omega}(r)}}\right) \eqcom \bar{\omega}(r) = \dfrac{\omega_\delta(r)+ \alpha^2\omega_\psi(r)}{1+\alpha^2} \eqcom
\end{equation}
where $r = |x_1-x_2|$ and $T(z,a)$ is the T-Owen function defined as \cite{Owen1956}
\begin{equation}
T(z,a) \equiv \dfrac{1}{2\pi} \int_0^a \dd{t}\dfrac{e^{-\frac{(1+t^2)z^2}{2}}}{1+t^2} \eqdot
\end{equation}
For $a=1$, a simpler form can be found \cite{Owen2}
\be
T(z,1)=\frac{1}{8}\erfc\left(-\frac{z}{\sqrt{2}}\right)\erfc\left(\frac{z}{\sqrt{2}}\right) \ .
\ee

\subsection{Nearest neighbours distance distributions}\label{app:nearest_neighbour}
Here we follow the notation and approach of Refs. \cite{torquato1986,torquato1990,truskett1998density}. For a homogeneous distribution, the $N$-point density is given by
\begin{equation}
    \rho_N(x_1,\dots,x_N) = \rho^N g_{N}(x_1,\dots,x_N)  \,
\end{equation} 
where $g_{N}(x_1,\dots,x_N)$ is the $N$-point correlation function. For the homogeneous system $\rho_1(x_1) = \rho = \text{constant}$.

We are interested in the following probability
\begin{itemize}
  \item[] $Q(x,y)\dd{x}\dd{y} \equiv$  given that there is a PBH at some position (could be the origin), the probability that the nearest PBH lies at a distance between $x$ and $x+\dd{x}$ and the second nearest PBH lies at a distance between $y$ and $y+\dd{y}$.
\end{itemize}
To compute it, we need to define some quantities.
\begin{itemize}
  \item[] $H_n(r)\dd{r} \equiv $ given that there is a PBH at the origin, the probability that the $n$-th nearest PBH lies at a distance between $r$ and $r+\dd{r}$
\end{itemize}
for $n \geq 1$. Clearly, from the above definition we have
\begin{equation}\label{H2vsQ}
H_2(y) = \int_{0}^y Q(x,y)\dd{x}  \ .
\end{equation}
So, if we manage to express $H_2(y)$ as Eq. \eqref{H2vsQ} we can obtain an expression for $Q(x,y)$.

We define the following regions:
\begin{itemize}
  \item[] $\Omega(r) \equiv $ the volume of a sphere of radius $r$ encompasing the reference PBH.
  \item[] $s(r)\dd{r} \equiv$ the volume of the spherical shell of a sphere of radius $r$. 
\end{itemize}

Let us now introduce more quantities
\begin{itemize}
  \item[] $E_n(r) \equiv $ given that there is a PBH at some position (the origin), the probability that the region $\Omega(r)$, encompassing the central PBH, contains $n$ additional PBHs.
  \item[] $\rho s(r) G_n(r)\dd{r} \equiv$ given that there are $n$ PBHs in the region $\Omega(r)$ (in addition to the central PBH), the probability that PBHs are contained in the shell $s(r)\dd{r}$ surrounding the central PBH. 
\end{itemize}
The function $G_n(r)$ is a conditional pair correlation function.  Note that if all the correlation function can be expressed as products of the 2-point, then $G_n(r) = g_2(r)$, with $g_2(r)$ the pair correlation function (denoted by $G(r)$ in Ref. \cite{Ballesteros:2018swv}). By the above definitions, we can write then
\begin{equation}\label{Hprod}
H_{n}(r)\dd{r} =\rho s(r) G_{n-1}(r) E_{n-1}(r)\dd{r} \ .
\end{equation}
Moreover, $H_n(r)\dd{r}$ and $E_n(r)$ are related by
\begin{equation}\label{HvsE}
H_n(r) \dd{r} = - \sum_{i=0}^{n-1} \dfrac{\partial E_i(r)}{\partial r} \dd{r} \ \text{, or} \sum_{i=0}^{n-1}E_i(r) = 1 - \int_0^r H_n(r')\dd{r'} \ .
\end{equation}
Let us first find an expression for $H_1(x)$. By using Eq. \eqref{H2vsQ} we can write Eq. \eqref{Hprod} as
\begin{equation}\label{eq:b5}
- \dfrac{\partial E_0(x)}{\partial x} =\rho s(x) G_0(x) E_0(x) \implies E_0(x) = \exp\left(- \int_{0}^x \rho s(x') G_0(x')\dd{x'}\right) \ .
\end{equation}
The lower bound acually should be $R_{\text{BH}}$, but we can set it later. The lower limit is set by imposing the condition that $E(0)$ (or $E(R_{\text{BH}})$) is one, since for sure there will be no PBHs (apart from the central one). Then we can write $H_1(x)$ as
\begin{equation}\label{h1}
H_1(x)\dd{x} = \rho s(x) G_0(x) \exp\left(-\int_0^x \rho s(x') G_0(x')\dd{x'}\right) \dd{x}
\end{equation} 
This is normalized, no matter the lower bound. Indeed if we consider the variables $X = \int_0^x \rho s(x') G_0(x')\dd{x'} $ then
\begin{equation}
H_1(X)\dd{X} = \exp(-X)\dd{X} \ , \ X \in (0,\infty) 
\end{equation}
Let us now compute $H_2(y)$ in a similar way. Note that
\begin{equation}
H_2(y) = H_1(y) - \dfrac{\partial E_1(y)}{\partial y}
\end{equation} 
Then we can write Eq. \eqref{Hprod} as
\begin{equation}\label{e1ode}
 \dfrac{\partial E_1(y)}{\partial y} + \rho s(y) G_1(y) E_1(y) = H_1(y)
\end{equation}
This is a first order ODE for $E_1(y)$ and can be solved by the use of the integrating factor. Consider the integrating factor
\begin{equation}
I(y) = \exp\left(\int_0^{y} \rho s(z) G_1(z)\dd{z}\right) 
\end{equation}
and multiply Eq. \eqref{e1ode} by $I(y)$. Then we can write it as
\begin{equation}
 \dfrac{\partial I(y) E_1(y)}{\partial y} = I(y) H_1(y)
\end{equation}
So  
\begin{equation}\label{eq:b12}
E_1(y) = \exp\left(-\int_0^y \rho s(z) G_1(z) \dd{z} \right) \left( \int_0^y  H_1(x)\exp\left(\int_0^x \rho s(z) G_1(z) \dd{z}\right)\dd{x} + C \right)
\end{equation}
where $C$ is a constant to be determined. Observe that now we have that $E_1(0) = 0$ (or $E_1(R_{\text{BH}})=0$) since the probability of having one PBH inside the volume $\Omega(0)$ (or $\Omega(R_{\text{BH}})$) is zero. Then we require $C=0$. Hence we can write Eq. \eqref{Hprod} as
\begin{equation}
H_2(y) = \int_0^y  \rho^2 s(x) s(y) G_0(x) G_1(y) \exp\left[-\left(\int_0^x \rho s(z) G_0(z) \dd{z} +\int_x^y \rho s(z) G_1(z) \dd{z}\right)\right]\dd{x} 
\end{equation} 
where we used Eq. \eqref{h1}, we put the $\rho s(y) G_1(y)e^{-\int^y \rho s(y) G_1}$ factor inside the $x$ integral and used the fact that $x<y$. Therefore we can write $Q(x,y)$ as
\begin{equation}
Q(x,y) =\rho^2 s(x)s(y)  G_0(x) G_1(y) \exp\left[-\left(\int_0^x \rho s(z) G_0(z) \dd{z} +\int_x^y \rho s(z) G_1(z) \dd{z}\right)\right] \Theta(y-x) \ .
\end{equation} 
$G_n$ will depend on all the $N$-point correlation functions $g_N$. Indeed we can write $E_n$ as \cite{truskett1998density}
\be\label{eq:En_corr}
E_n(r)=\frac{1}{n!}\left[\left(\dfrac{\partial }{\partial t} \right)^n \left( 1+\sum_{i=1}^{N-1}\dfrac{t^i}{i!}\rho^i \int g_{i+1}\left(\mathbf{r}_{12},...,\mathbf{r}_
{1i}\right)\prod_{k=2}^{i+1}\Theta\left(r-|\mathbf{r}_{1k}|\right)\dd \mathbf{r}_{1k} \right) \right]_{t=-1}  \ ,
\ee
where $\mathbf{r}_{1i}=\mathbf{r}_1-\mathbf{r}_i$. If the separability condition \eqref{eq:sep} holds, then
\begin{align} \label{eq:eo}
 E_0(r)&=\exp\left( - \int \rho s(z) g_2(z) \dd z  \right) \ , \\ \label{eq:e1} 
 E_1(r)&=  \left[ \int \rho s(z) g_2(z) \right]\exp\left(-\int\rho s(z) g_2(z) \dd z  \right) \ ,
\end{align}
which combined with Eqs. \eqref{eq:b5},\eqref{h1} and \eqref{eq:b12} give
\be
G_0(r)=G_1(r)=g_2(r) \ .
\ee

\bibliographystyle{plain}

\end{document}